# Review of studies in field of the effects of nanotechnology on breast cancer

ª*PARISA TEIMOORI BAGHAEE, ᵇATENA DONYA

ª,ᵇ*Islamic Azad University Mashhad Branch, Mashhad, Iran*
ª *Parisa.t89@gmail.com*

Abstract: Cancer is the main cause of mortality at the developed countries and is the second cause of mortality at the developing countries and breast cancer is the most prevalent malignancy and the first cause of mortality among women of the world. At the US, 28% of new cancer cases and 15% of mortality caused by that in 2010 was caused by breast cancer. Moreover, breast cancer is the most common cancer among Iranian women, which includes 24.4% of overall malignancies and causes 3.3% of mortalities per 1.000 people. The disease is most spread in Tehran and including 25.5% of all cancers. According to the mentioned, the study has tried to assess the effect of nanotechnology in form of a review to diagnose and treat breast cancer. The method applied in this study is descriptive analytical method and library method has been used for data collection purpose. In the data collection method, 45 articles relevant to structure of types of nanoparticles and their uses in diagnosis, imaging and medicine delivery systems and breast cancer treatment have been used. Although there are still some challenges and limitations to use nanoparticles in medication, it is hope that nanoparticles can make wonderful revolution not only in oncology, but also in medication in near future. The most underlying headlines of the present study include mineral nanoparticles, antibodies and tumor imaging methods.

## 1 Introduction

The nanotechnology science has helped diagnosis and treatment of diseases by using nanoparticles and development of systematic drug delivery methods. Nanoparticles have high potential in field of diagnosis and treatment of diseases, especially cancer. Using these nanoparticles as factors increasing contrast in conventional imaging method of magnetic resonance and as nano-carriers in modern drug delivery systems have gained attention of scholars over the years. Systematic transfer of chemotherapy factors to cancer cells using magnetic nanoparticles under in vivo and in situ conditions has been studied and valuable results have been obtained. Moreover, the nanoparticles are applicable in treatment of cancer using heating method and transfer of nucleic acids, plasmids and siRNA to cells. Diagnosis of the disease in early stages can be underlying for improvement and treatment methods. recently, the treatment methods used to diagnose cancer are taken usually based on changes in cells and tissues and this can be also done by medical clinical examinations and using conventional imaging methods. However, scientists tend to diagnose cancer with appearance of the first molecular changes. Iron oxide nanoparticles are currently the only nanomaterials used in clinical medicine as the factor for contradiction in magnetic resonance imaging and as carrier in drug delivery. Experiments taken over the years on iron oxide nanoparticles show that the particles have no immediate toxic or long-term effect in vivo conditions; although presence of some nanoparticles, along with nano-carriers, can enhance their effect and function on cancer cells. To this end, the nanoparticles can reinforce nano-carriers by means of different mechanisms such as increased oxidative stress and proper drug accumulation in the cell [1-2]. Annually, more than 1.1million cases of breast cancer are being reported and more than 410.000 people die because of breast cancer [2]. There are many treatment methods for breast cancer such as surgery, radiotherapy and hormone therapy, which have lots of complications and disadvantages. For example, tamoxifen used in chemotherapy for treatment of breast cancer can cause cancer in endometrial tissues. Therefore, it is required to make a system to deliver drug to tumor tissue with zero complications [1]. Using nanoparticles in food industry, electronics and medication is being developed. Preparing particles in nano size (smaller than 100nm) has increased their levels and has increased their reaction with organic and nonorganic molecules [3]. Using nanotechnology in medication refers to use very tiny articles (1-1000nm) in field of diagnosis and treatment of human cancers causing a new scientific branch in oncology under the title of nano-oncology [4, 5]. Nanoparticles have been developed for imaging, showing biomolecules and cancer biomarkers and systemizing drug carrying. Fixing chemotherapy drugs, especially in kind of enzyme in polymer nanoparticles, has led to increase their resistance to heat, pH, proteases and other destructive factors [6]. Semi-Conductor Fluorescent Nonocrystals such as quantum dots conjugated with antibodies can cause marking determining their exact amount in a small piece of breast tumor [7]. Other particles in nano scale such as nano-cantilerer and nano probes and coupled nanoparticles with specific ligands have been studied in field of cancer tumor imaging and determining environmental metastasis [8, 9]. Relevant studies have shown that nanoparticles conjugated with metal core and super magnetic force with biologic antibodies against ERBB2 gene can simultaneously be useful in imaging and treatment of breast cancer [10]. Over the years, advancement of nanotechnology has caused use of nanoparticles in different fields of electronics, optic devices, industries, disease diagnosis, drug delivery, biosensors, imaging, and production of consumptive products such as sunscreens, cosmetics and sport instruments. The aim by this study is to investigate the studies in field of structure of types of nanoparticles and using them in diagnosis, imaging and drug delivery systems and to treat cancer.

## 2 Methodology

For purpose of data collection, databases such as Pubmed and Google Scholar were used. The researches and reviews using MeSH model included malignancy, cancer, neoplasm, nanotechnology, nanoparticles, aptamers, tumor imaging, magnetic nanoparticles, liposomes, biomarkers, interference small RNA, diagnosis and treatment methods of breast cancer using nanoparticles and nanobodies.

## 2.1 Types of nanoparticles used in medication

The nanoparticles used in medication field are classified in 2 main groups: particles containing organic molecules as the main structural substance and second group include metals and minerals as central core. Liposomes, dendrimers, carbon nanotubes (CNTs), emulsions, aptamers, solid lipid nanoparticles, nanobodies and other polymers have been recognized as organic particles [11-13].

## 2.2 Liposomes and dendrimers

Liposomes can be used as chemotherapy drug carrier for different human tumors such as breast cancer [14] (figure 1). Dendrimers can be used as contrast factor in MRI and can be the assisting factor to clear pathologic procedures [15, 16].

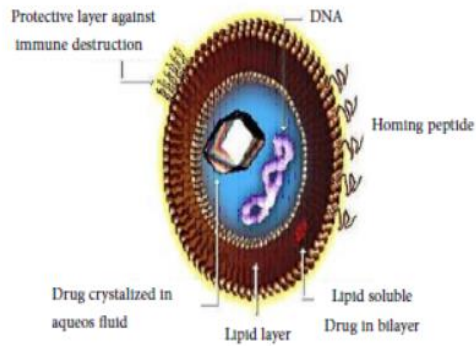

Figure 1: drug-carrying liposome [14]

## 2.3 Aptamers

Aptamers are oligonucleotide (RNA or ssDNA) and peptides molecules attached to target molecule (small biomolecules, proteins and even cells) with high tendency and specificity and can be instruments to diagnose and treat diseases such as cancer [13]. DNA stability of aptamers is higher than RNA; although RNA aptamers have higher flexibility.

Peptide aptamers have been made of a variable peptide loop attached in the end to a protein scaffold. The variable loop includes 10-20 amino acids and protein scaffold incudes any kind of protein with high solubility. Moreover, aptamers can be used in drug carrying systems. The aptamers have been bonded to cell surface receptors and have trend to be dragged inside the cell. siRNAs can be used todat as new technique for treatment. Their function is induction of iRNA to regulate special gene expression. Safe, specific and efficient delivery of siRNAs inside the special cells is underlying in field of treatment and medication [17].

## 2.4 Nanobodies

Sharks have antibodies without light-chain, and contain just heavy-chain and are called heavy chain antibodies (HCAbs). The antibodies lack CH1 domain; although they contain CH2 and CH3 domains, which are highly similar to common antibodies. Therefore, Fb part bonded to antigen in conventional antibodies is limited to a variable domain in HCAb antibodies. As it was mentioned, it includes just heavy chain called HHV. As its size is in limit of nano scale (2-5nm in diameter and 4nm in height), they are called nanobodies and can be used abundantly in nanotechnology. Nowadays, monoclonal nanobodies can be produced in bacteria and they are more resistant than conventional antibodies because of their small size, have high solubility and have high specificity to antigen. Therefore, they have high potential to diagnose and treat cancer [2].

In the studies conducted by Katrien Van Impe et al on rats, using nanobodies could decrease the metastasis of cancer cells of breast glands in these rats [18]. In 20-30% of breast cancers, it has been observed that the expression and activity of 2 surface antigens of HER1 and HER2 was increased [19]. Through making specific nanobodies of the two surface receptors, cancer can be diagnosed and treated.

## 3. Findings

### 3.1 Magnetic nanoparticles, diagnosis and treatment

The issue of systematic drug carrying is mostly considered for cancer treatment, since the main challenge in treatment of cancer is targeting and destroying cancer cells, so that it can have lowest effect on healthy cells. One of the main purposes of nanotechnology is carrying drugs on carriers (nanoparticles), sending, and leaving them in target cells, which can be called as systematic drug delivery. Using magnetic nanoparticles and creating a magnetic field, drug can be delivered to desired tissue intelligently and cause healing in tissue without damaging other tissues [20, 21]. In order to use magnetic nanoparticles in medical uses, those showing super paramagnetic phenomenon in ambient temperature are preferred, since the magnetic property of these particles can be lost in absence of magnetic field and cannot remain. Moreover, existence of magnetic remains in these materials can lead to their coagulation and this can cause blockage of blood in vessels. Therefore, for such uses, using compounds with super paramagnetic property is essential [21-25]. In general, super paramagnetic property is a kind of magnetic property observed in ferromagnetic small nanoparticles. In sufficiently tiny particles, magnetism can be changed randomly under the effect of temperature. In absence of external magnetic field, when the time sued to measure magnetism of nanoparticles is significantly longer than rest time, average magnetism of the particles becomes zero and it could be mentioned that particles are under super paramagnetic conditions (figure 2).

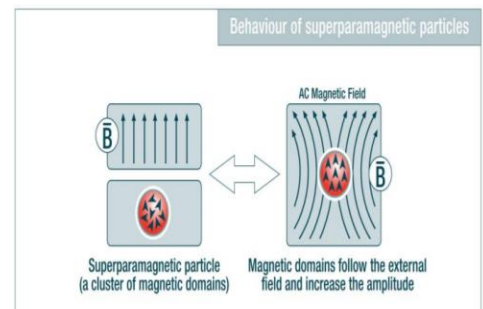

Figure 2: behavior of super paramagnetic particles applicable in biomedical fields, especially accumulation of nanoparticles in place of tumor [9]

Underlying for medical diagnosis uses such as nanoparticles to enhance the contrast of magnetic resonance images [33]. Underlying points about nanoparticles used in medical cases are that the size and distribution of size of particles should be thin. Magnetic nanoparticles should have desirable form, size and surface properties to shoot the tumors. For example, in biologic uses and medical diagnosis, magnetic nanoparticles can be used without cell toxicity and stable in water and natural pH and human body physiologic conditions. With taking required modifications on surface of these compounds, the conditions can be provided to bond the drug, protein and genetic materials (figure 3). Hence, targeted transfer of these compounds can be achieved with the effect of an external magnetic field [34-35].

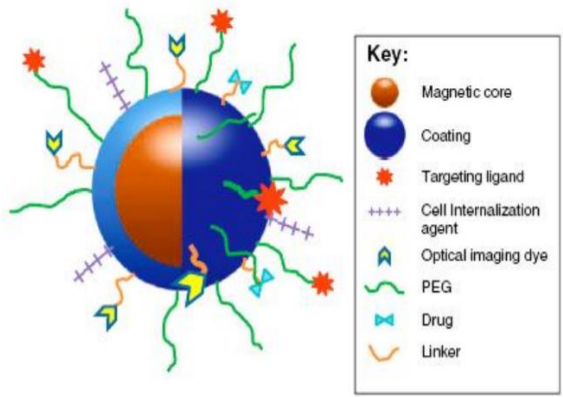

Figure 3: schematic of shell/core structure of magnetic nanoparticles and decoration of their surface with different functional groups.

These particles have been formed of magnetic iron oxide core and a coating of biocompatible materials such as polysaccharides, Dextran, Chitosan, Polyethylene glycol and polyvinyl alcohol. Functional groups in coating surface of nanoparticles have been used to link the ligands, molecular targeting, transferring nanoparticles to cancer cells and systematic drug delivery and other medical uses such as hyperthermia and magnetic resonance imaging [12].

### 3.2 Tumor imaging (diagnosis)

Currently, magnetic nanoparticles are being used because of their capability in making contrast in MRI [36], which has lower toxicity and biologic adaptability [37]. Super paramagnetic nanoparticles (0-10nm) can be used as contrast factor in MRI and bismuth nanoparticles can be used as contrast factor for CT. despite to their good absorption, they have nonspecific distribution and are fast in terms of pharmacokinetics. If the surface of the bismuth nanoparticles is covered by polymer, they can be protected from decomposition and the tissues can be protected against their toxic effect. The nanoparticles have shown good stability in high concentration, they can absorb high range of x-ray and their survival time in blood reaches to more than 2 hours. Therefore, the efficiency ratio of these particles to their non-toxicity is better than iodized materials in imaging [36]. Using quantum particles with infrared wavelength has been also reported for non-invasive tumor imaging in vivo [37].

Gao et al [9] covered quantum particles using triplet copolymer amphiphilic (poly-methacrylate hydrophilic components and two poly-butyl acrylate hydrophobic parts and a polyethylene acrylate unit). The polymers protect nanoparticle against decomposition and as they contain polyethylene glycol molecules, they can improve biologic ability of quantum particles and their intercellular circulation. Conjugation of quantum particles with systematic antibodies against membrane specific genes of prostate can mark the human prostate cancer and can decrease the accumulation of quantum particles in liver and bone marrow. Moreover, 3 polymers with dimensions of 5.0μm of quantum particles with green, yellow and red colors can be observed simultaneously in 3 different parts. Despite to dextran conjugated with organic colors and influenced in tumor vessels and expose them to watch transparently; quantum particles mark vascular wall clearly, which can show expression of proteins containing green fluorescent color. Quantum particles link to endothelial sensitive tissue and target the lymph vessels around the tumor. Using quantum points with distribution range near the infrared have been recommended for tumor imaging. Determination of lymph nodes can be used to specify degree of melanoma and breast cancer. Conventional method is using blue color or injection of radioisotopes. Today, quantum nanoparticles have been also recommended to determine degree of cancer and for early diagnosis of cancer. The limitation of using quantum particles is because of their toxicity and this method is being developed with reduction of toxic effects of particles [38, 39].

### 3.3 Breast cancer treatment with magnetic hyperthermia processes

Cancer treatment is one of the most important challenges for medical science and drug delivery. As the conventional treatments such as surgery, chemotherapy and radiotherapy are not so effective to treat some cancers such as glioblastoma; hyperthermia can be considered as a reliable approach for absolute treatment of these cancers. Cancer treatment by means of inductive irritation of super paramagnetic and biocompatible nanoparticles by means of alternative magnetic field of heating special tissues or organs to temperatures about 41-47°C for cancer treatment can be called hyperthermia (figure 4).

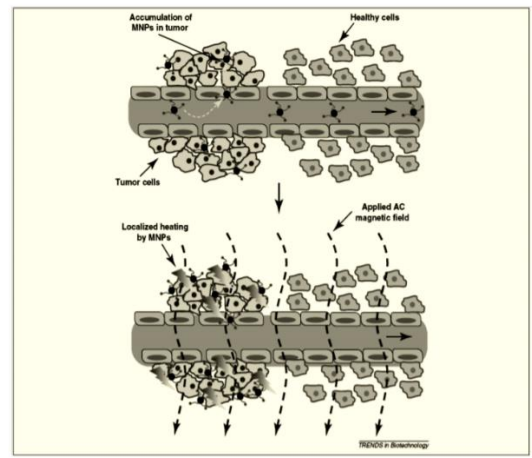

Figure 4: a schematic of magnetic hyperthermia principles; the principles have been formed based on accumulation of magnetic nanoparticles in tumors compared to healthy tissues [12]

Hyperthermia process with magnetic ferrofluid suggests possibility of positioning heat. This is because; cancer cells are sensitive to overheat of normal cells. Super paramagnetic nanoparticles can transmit the additional heat to target point through fluctuations of magnetic moments inside the nanoparticles. Therefore, solid cancer cells will be killed; although normal tissue cells remain under temperature below 41-47°C. Heating potential is highly depended on form and size of particles. Therefore, using single-organized magnetic particles in nano scale is prior to multi-organized particles in micro scale. This is because; nanoparticles can tolerate alternative magnetic field. The produced heat varies depending on frequency, magnetic field resonance and time of placement in magnetic field. Under this conditions, after placement of nanoparticles inside cells and placement of body under magnetic field; the particles produce positional and specific heat inside the cancer mass and accelerate destruction of tumor. Hence, hyperthermia can enhance treatment solid cancer [12].

### 3.4 Simultaneous imaging and treatment

Nanoparticles conjugated with targeted antibodies are capable to be used simultaneously for diagnosis and treatment of cancer. Conjugation of target ligands can be done using Biotin and Streptovidine. This method can be used to conjugate an ERBB2 antibody to a metal nanoparticle to form a nanoshell [40]. The structure of this system is formed of a rounded nanoparticle in dielectric core and silica and surrounded by thin layer of gold. The nanoshell with a transmission spectrum close to infrared convert light to thermal energy and takes function for surgery of tumor or removal of tumor using thermal method. the nanoparticles making heat under the effect of near-IR radiation have effective efficiency to 1million times higher than colored molecules. The heat created by them is higher than the amount needed to create irreversible tissue damage and can cause cell death ultimately. Targeted nanoshells can be used for irreversible removal of breast tumor in vivo conditions.

### 3.5 Breast cancer treatment

Wide range studies have been taken to use nanoparticles to protect healthy tissues and to have killing effects on cancer cells. For example, Liposomal anthracycline can be used to treat all degrees of breast cancer [41]. However, using that is restricted because of toxic effects on heart. The compound, along with Trastuzumab as a monoclonal antibody against ERBB2, has better effect [42].

It has been reported that liposome and Epiglottis liposome containing Doxorubicin can be used in treatment of metastatic breast cancer [43, 44]. Moreover, nanoparticles in core section containing paclitaxel and surrounded by albumin can be effective to carry hydrophobic molecules in breast cancer. Preclinical studies have revealed that paclitaxel contained in nanoparticles surrounded by albumin has better power of infiltration in tumor compared to conventional paclitaxel. Targeted delivery of tamoxifen can be taken in all stages of breast cancer. Tamoxifen is a Nonsteroidal and anti-estrogenic drug with high hydrophobic property [44]. Using nanoparticles with the drug can increase its influence to tumor. Moreover, its toxic effects on healthy tissues are low [45].

### 4 Discussion

As use of conventional diagnosis and imaging methods of tumor have some disadvantages and limitations; using nanotechnology like quantum nanoparticles, nanobodies and aptamers can be novel method for early diagnosis of cancer, especially breast cancer. In addition to diagnosis of cancer, aptamers can be effective for cancer treatment. Aptamer technology has been developed as a reliable and effective technology, many studies have been conducted on uses of aptamers, and the aptamers are being used in different dimensions as a diagnosis and treatment instrument and drug carrying systems. However, aptamers have not still found their position in field of medical diagnosis and treatment.

Nanoparticles conjugated with antibodies can be used simultaneously to detect various molecular targets in small tumor pieces. Moreover, with regard to side effects of anticancer drugs, using safe drug delivery systems with biocompatibility like solid lipid nanoparticles and liposomes delivering drug to target tissue with high specificity is essential. The basis of this method is making enough amount of drug reach to tumor place for certain time and to reduce harmful effects of drug on other organs. Solid lipid nanoparticles can be considered as a complicated system with unique advantages and disadvantages, which can separate it from other colloid systems. Further studies are needed using magnetic-resonance methods to clear drug delivery mechanism using the system. Limited information is available about toxicity aptamers. Therefore, to take aptamer therapy, it is necessary to know toxicity level. The main challenge in chemotherapy of cancer is unwanted effect and complications of drugs. The single dose or short term of using these drugs can cause serious risks for human health; although using biodegradable particles in nano scale for long term or treatment periods can cause unwanted harmful effects. Therefore, there are still some challenges and limitations to use nanoparticles in medication. It is hope that synthesis costs are reduced in near future and pharmacokinetic properties are increased to enhance productivity of nanoparticles and to overcome limitations of using them.

Because of special magnetic properties, magnetic nanoparticles have gained clinical importance for diagnosis and treatment cancer. With analysis of various studies conducted in this field, it could be found that loading anti-cancer drugs on magnetic nanoparticles can play effective role in enhancement of drug mechanism and removal of cancer cells. The research results have shown that magnetic nanoparticles can improve performance of anticancer drugs (Doxorubicin and Sysplatin) in killing cancer cells through improvement of production of active oxygen species or other unknown mechanisms. In other words, magnetic nanoparticles can increase cell toxicity of anticancer drug and plays also key role in transfer of drug to tumor cells. Magnetic resonance imaging is a common method used for diagnosis of tumor and cancer. Using magnetic nanoparticles in this method can cause sensitiveness o this method for cancer diagnosis, so that cancer can be diagnosed in early stages and effective treatment measures can be taken for its certain treatment. In addition, special magnetic property of magnetic nanoparticles can be used in targeted drug delivery. The nanoparticles, especially iron oxide magnetic nanoparticles, can be used as nanocarrier to carry the drug because of capability of encapsulation of anticancer drugs and high biocompatibility and low biologic toxicity and high concentration capability in tumor. Through this, the side effects of the drug can be decreased significantly. According to the relevant studies and results obtained from this study, it could be mentioned that modern drug carrying and multifunctional systems can be designed in near future with high capability to carry drug to tumor point. In addition, they can be simultaneously used as contrast factor in magnetic resonance imaging and magnetic hyperthermia process for certain treatment of cancer.

### 5 Conclusion

In this study, in regard with nanoparticles, the advantages and uses of these particles for treatment and diagnosis of breast cancer are discussed. Nanoparticles can be used in medication in field of diagnosis and treatment of diseases, drug carrying and biologic imaging. Advancement of nanotechnology in oncology can provide special facilities to identify the various molecular targets simultaneously in small tumor samples to take treatment strategy. Using nanoparticles in the in vivo tumor imaging is being developed rapidly and targeting antigens relevant to cancer can be possible in future. In near future, nanotechnology science can make wonderful revolution not only in oncology, but also in all steps of medical sciences.

Molecular targeting has been studied in field of development and advancement of tumor selecting factors used in magnetic resonance imaging. The factors promise more medical uses. Macromolecular ligands have gained high attractiveness to be used in targeting nanoparticles; although expensiveness of synthesis of these ligands and complicated nature of chemistry of bonding them to magnetic nanoparticles has made some problems and limitations in regard with using these factors in clinical and laboratory phases. Moreover, extension of cation capabilities to magnetic nanoparticles by peptides influencing in cell can shorten their residence in plasma and can remove them. Therefore, while using these materials to design platform and to make sure of selectiveness of tumor and for Pharmacokinetic preservation, it is required to make decisions to adjust cation activity. Any kind of proof on nanoparticles and clearing their functional mechanisms can be helpful to design novel anticancer nanocarriers. The nano-carriers allow decreased use of cytotoxic drug with their mechanism and allow also overcoming resistance of cancer cells to drug.